# Size Dependence of Lattice Parameter and Electronic Structure in CeO$_2$ Nanoparticles

*Damien Prieur*[1,2*], *Walter Bonani*[3,*], *Karin Popa*[3], *Olaf Walter*[3], *Kyle W. Kriegsman*[4], *Mark H. Engelhard*[5], *Xiaofeng Guo*[4], *Rachel Eloirdi*[3], *Thomas Gouder*[3], *Aaron Beck*[6], *Tonya Vitova*[6], *Andreas C. Scheinost*[1,2], *Kristina Kvashnina*[1,2], *Philippe Martin*[7]

[1] Helmholtz Zentrum Dresden-Rossendorf (HZDR), Institute of Resource Ecology, PO Box 510119, 01314 Dresden, Germany.

[2] The Rossendorf Beamline at ESRF—The European Synchrotron, CS40220, 38043 Grenoble Cedex 9, France

[3] European Commission, Joint Research Centre, P.O. Box 2340, D-76125 Karlsruhe, Germany.

[4] Department of Chemistry and Alexandra Navrotsky Institute for Experimental Thermodynamics, Washington State University, Pullman, WA 99164

[5] Environmental Molecular Sciences Laboratory, Pacific Northwest National Lab, Richland, WA 99352

[6] Institute for Nuclear Waste Disposal, Karlsruhe Institute of Technology, P.O. Box 3640, 76021 Karlsruhe, Germany.

[7] CEA, DEN, DMRC, SFMA, LCC, F30207 Bagnols sur Cèze cedex, France.

**KEYWORDS** Lanthanide, CeO$_2$, HEFRD-XANES, Electronic Structure

**ABSTRACT**

Intrinsic properties of a compound (e.g. electronic structure, crystallographic structure, optical and magnetic properties) define notably its chemical and physical behavior. In the case of nanomaterials, these fundamental properties depend on the occurrence of quantum mechanical size effects and on the considerable increase of the surface to bulk ratio. However, the literature on this size-dependence and on the involved mechanisms is quite elusive and scarce. Here, we explore the size-dependence of both crystal and electronic properties of $CeO_2$ nanoparticles (NPs) with different sizes by state-of-the art spectroscopic techniques. XRD, XPS and HERFD-XANES demonstrate that the as-synthesized NPs crystallize in the fluorite structure and they are predominantly composed of $Ce^{IV}$ ions. The strong dependence of the lattice parameter with the NPs size was attributed to the presence of adsorbed species at the NPs surface thanks to FTIR and TGA measurements. In addition, the size-dependence of the $t_{2g}$ level in the Ce $L_{III}$ XANES spectra was experimentally observed by HERFD-XANES and confirmed by theoretical calculations.

**INTRODUCTION**

$CeO_2$-based nanoparticles (NPs) offer unique redox properties that open promising possibilities for applications in catalysis,[1,2] energy storage,[3,4] biomedicine[5], and nuclear activities.[6] Quantum mechanical size effects, combined with a considerable increase of the surface to bulk ratio, are responsible for the unique properties of nanometer-sized particles, including electronic and geometric structure, and optical and magnetic properties.[7,8] A thorough understanding of the dependence of these properties on particle size is of great importance not only for the design of next generation materials.

In this context, our work focuses on studying the size-dependence of both crystallographic and electronic structures of $CeO_2$ NPs as these two fundamental properties are of technological importance and theoretical interest and broad prospects.[2] The change of unit cell dimensions with decreasing particle size has been previously reported, but remains a subject of discussion. Different hypotheses have been put forward to explain this phenomenon: surface stress induced by the presence of sorbed species and partial reduction of $Ce^{IV}$ to $Ce^{III}$.[9–11] Furthermore, the electronic structure and its size-dependence, is of fundamental interest as energetic and catalytic properties notably lie on it.[2]

In the present work, the crystal structure of as-synthesized $CeO_2$ NPs was characterized by using x-ray diffraction (XRD) and transmission electron microscopy (TEM) giving access to their size and lattice parameter. Fourier transform infrared spectroscopy (FTIR) and thermogravimetric analysis (TGA) was performed to determine species potentially absorbed at the surface. Furthermore, we probed the electronic structure of $CeO_2$ NPs using high-energy resolution fluorescence-detection hard X-ray absorption near-edge structure (HERFD-XANES) spectroscopy at the Ce $L_{III}$ edge. Thanks to the use of an X-ray emission spectrometer, such inner-shell spectroscopy provides an element-selective probe of the electronic state and allows observing spectral features with significantly enhanced energy resolution compared to usual data limited by Ce $L_{III}$ edge core hole life-time broadening (Supporting information Figure 1) [12,13].

**EXPERIMENTAL METHODS**

**Synthesis method**

   **Nanoparticles obtained by hydrothermal treatment**

Ceria nanoparticles samples were synthesized by alkaline precipitation of cerium ammonium sulfate precursor followed by hydrothermal condensation in a pressurized autoclave at different temperatures[14,15]. In detail, a 1M Ce(IV) solution was prepared by dissolving cerium(IV) ammonium sulfate dehydrate (Alfa Aesar) in deionized water; cerium(IV) hydroxide was directly precipitated by adding an excess of ammonium hydroxide (Sigma Aldrich, 25% in water) under constant stirring for 3 hours. A yellow cerium(IV) hydroxide precipitate was recovered by centrifugation, repeatedly washed with deionized water and hydrothermally treated in a stainless steel reactor vessel with Teflon insert (total free volume 12 ml). Typically, 200 mg of cerium(IV) hydroxide were suspended in 10 ml deionized water and heated for 3 h under autogenous pressure at different temperatures. After cooling, the solid residue was recovered, washed with deionized water, dehydrated with ethanol and acetone and dried overnight in a chemical fume hood. The resulting dry powders were analyzed by XRD with a Rigaku Miniflex 600 diffractometer. The crystallite size of the nanopowders was estimated from the XRD pattern using the Scherrer equation and averaging the results of 8 selected peaks in the 2θ range between 25 and 80°. Nanoparticle samples obtained by hydrothermal condensation at 150, 180 and 200°C were labelled as Ce_2, Ce_4.2 and Ce_5.6, giving their respective crystallite size in nanometer.

**Nanoparticles obtained by thermal treatment in dry conditions.**

In order to obtain samples with larger crystallites size, the dry powder samples obtained by hydrothermal treatment were calcined for 1 h at temperatures ranging from 350 to 950 °C in an open furnace. The so obtained nanoparticle samples were analyzed by XRD and labelled on the basis of the respective estimated crystallite size. The nomenclature and the synthesis route of the different nanoparticle samples are reported in detail in Table 1.

Considering that the sample Ce_2 experienced damages due to the exposure to the beam during the HERFD-XANES measurement, this compound has been discarded from this discussion on the oxidation state determination and on the electronic structure. Further details on this phenomenon are provided in the Supporting information Figure 2.

**Transmission Electron Microscopy**

Transmission electron microscopy (TEM) studies were performed using an aberration (image) corrected FEI™ Titan 80-300 operated at 300 kV providing a nominal information limit of 0.8 Å in TEM mode and a resolution of 1.4 Å in STEM mode. TEM micrographs have been recorded using a Gatan US1000 slowscan CCD camera, while STEM images have been recorded using a Fischione high-angle annular dark-field (HAADF) detector with a camera length of 195 mm. The samples for analysis have been prepared by dropping coating with a suspension of the nanoparticles in ultrapure water on carbon coated copper grids.

**Fourier transform infrared spectroscopy**

Dehydrated ceria NPs were analysed by FTIR in attenuated total reflectance mode with an Alpha Platinum Bruker spectrometer equipped with ZnSe crystal. FTIR spectra were were obtained at room temperature in the wavenumber range from 600 to 4000 $cm^{-1}$ with a resolution of 4 $cm^{-1}$.

**Thermogravimetric analysis**

The thermal behaviour of ceria NPs was investigated using a Netzsch STA 449C DTA/TG using an alumina crucible and in air atmosphere. The temperature was controlled by a Pt-PtRh (10 %) thermocouple. Measurements were carried out at a constant heating and cooling rates of 10 ºC/min between 40 and 700°C.

**Raman**

Raman measurements of nanocrystalline samples were performed at room temperature with a Horiba Jobin-Yvon T64000 spectrometer using a Kr+ laser with excitation wavelength of 647 nm. A 50x objective was used to irradiate powder samples and collect the back-scattered light. The analyses were performed with an incident laser power in the 4-10 mW; no effect of laser power was observed for the resulting spectra in this range.

**X-ray Photoelectron Spectroscopy**

XPS measurements were performed with a Physical Electronics Quantera Scanning X-ray Microprobe. This system uses a focused monochromatic Al $K_\alpha$ X-ray (1486.7 eV) source for excitation and a spherical section analyzer. The instrument has a 32 element multichannel detection system. The X-ray beam is incident normal to the sample and the photoelectron detector is at 45° off-normal. High energy resolution spectra were collected using a pass-energy of 69.0 eV with a step size of 0.125 eV. For the Ag $3d_{5/2}$ line, these conditions produced a FWHM of 0.92 eV ± 0.05 eV. The binding energy (BE) scale is calibrated using the Cu $2p_{3/2}$ feature at 932.62 ± 0.05 eV and Au $4f_{7/2}$ at 83.96 ± 0.05 eV. The sample experienced variable degrees of charging. Low energy electrons at ~1 eV, 20µA and low energy Ar$^+$ ions were used to minimize this charging. The binding energy scale was charge corrected referencing the Ce $3d_{3/2}$ $4f^0$ (u''') line at 916.7 eV.[16,17] The quantification and peak fitting was performed using PHI MultiPak version 9.6.1.7 software, where the experimental results were fitted to 10 peaks, defined as $v^0$, v, v', v'', v''', $u^0$, u, u', u'', and u'''.[16] In this scheme, v, v'', v''', u, u'', and u''' were associated with $Ce^{IV}$, while $v^0$, v', $u^0$, and u' were associated with $Ce^{III}$. The relative areas of the fitted peaks were measured in order to determine the relative concentration in the surface.

**High-Energy Resolution Fluorescence-Detected X-ray Absorption Near Edge Structure**

HERDF-XANES measurements were conducted at the CAT-ACT beamline (ACT station) of the KIT synchrotron light source (Karlsruhe Institute of Technology, Karlsruhe, Germany).[18,19] The incident energy was selected using the (111) reflection of a double Si crystal monochromator. The X-ray beam was focused to 500 x 500 µm onto the sample. XANES spectra were measured in high-energy-resolution fluorescence detected (HERFD) mode using an X-ray emission spectrometer.[12,19] The sample, analyzer crystal, and a single diode VITUS Silicon Drift Detector (KETEK, Germany were arranged in a vertical Rowland geometry. The Ce HERFD-XAS spectra at the $L_{III}$ edge were obtained by recording the maximum intensity of the Ce $L_{α1}$ emission line (4839 eV) as a function of the incident energy. The emission energy was selected using the ⟨331⟩ reflection of four spherically bent Ge crystal analyser (with a bending radius R = 1 m) aligned at a 80.7° Bragg angle. The experimental energy resolution was 1.15 eV obtained by measuring the full width at half maximum of the elastically scattered incident beam with an energy of 4.8404 keV. During the measurements a slit with the dimensions 500 x 500 µm was used in front of the sample, cutting of tails in the pprofile of the incident beam. This potentially led to a slight improvement of the experimental resolution, i.e. 1 eV (it is an estimation). The experimental resolution measured at ID26, ESRF using the same analyzer crystals was 0.9 eV[20]. The sample, crystals and detector were confined in a box filled with He and a constant He flow was maintained in order to minimize the loss of intensity due to absorption and scattering of the X-rays. The data were not corrected for self-absorption effects. The sample exposure to the beam was minimized to account for possible beam damage and checked by first collecting short XANES scans (~10 seconds) to look for irradiation effect.

**Theoretical calculations**

The spectra of bulk $CeO_2$ and $CeO_2$ in 2nm were performed in a manner described in [21–24] using the FEFF 9.6. code. Similar to the work of Li *et al.* and Plakhova *et al.*, we show here only part of the absorption spectra, which corresponds to the 2p–5d transitions, and omit multi-electron excitations from the experimental data, which appear at higher incident energy.

**RESULTS AND DISCUSSION**

**Ce valence in the as-synthesized $CeO_2$ nanoparticles.**

Our X-ray diffraction (XRD) and transmission electron microscopy (TEM) data (Supporting Information Figure 3 and Figure 4) show that the as-synthesized $CeO_2$ NPs crystallize in the *Fm-3m* fluorite structure (*space group* 225). Depending on the experimental conditions (and particularly the annealing temperature), the average crystallite diameters vary from $2.0 \pm 0.1$ to $91 \pm 12$ nm. These XRD-refined parameters are gathered in the Table 1 and will be more thoroughly discussed in a following section.

<center><Table 1 about here></center>

The oxidation state and electronic structure of Ce was assessed using Ce $L_{III}$ edge HERFD-XANES and XPS, which corresponding spectra are respectively given in Figure 1 and in Figure 2.

<center><Figure 1 about here></center>

The HEFRD-XANES spectra of all the investigated $CeO_2$ NPs exhibit a single pre-edge peak aligned with that of the bulk-$CeO_2$ reference spectrum. This pre-edge peak (noted A in Figure 1) originates from the 2p transition to a mixed 5d-4f valence state [22,25] and is characteristic of the Ce valence of the probed sample. Indeed, a single peak is observed in the pre-edge region of a pure $Ce^{IV}$ compound, since the photo absorption process excite an electron to the 4f level, formally empty in the initial state. However, in the case of a $Ce^{III}$ ion, the interaction between the 4f electron in the initial state and the second electron excited by the photon leads to a splitting of the pre-edge feature into two groups of transition, whose energy position is related to the electron–electron interactions in the 4f level.[20,26,27] Consequently, the pre-edge structure of the NPs reveals solely the presence of Ce in the IV oxidation state. Additionally, XANES being extremely sensitive to the local structure, those similarities corroborate our XRD findings that $CeO_2$ NPs all crystallize in the same space group.[8]

<Figure 2 about here>

Regarding our XPS results, Figure 2 provides the Ce $3d_{3/2,5/2}$ XPS spectra collected for the $CeO_2$ NPs. The spectra of $CeO_2$ are composed of 6 peaks corresponding 3 doublets belonging to $3d_{3/2}$ and $3d_{5/2}$ core holes spin-orbit splitting. The highest energy peaks, u''' at 916.7 eV and v''' at 898.2 eV, are the doublet from the ejection with the Ce $3d_{3/2}$ $4f^0$ final state. The next doublet, u'' at 907.4 eV and v'' at 888.8 eV, are the result of Ce $3d_{5/2}$ $4f^1$, followed by u at 900.8 eV and v at 882.4 eV, corresponding to the result of a Ce $3d_{5/2}$ $4f^2$ final state. All of these multiplets are associated with $Ce^{IV}$, which are consistent with previously published $CeO_2$ reference XPS data. On the other hand, the Ce 3d XPS spectrum of $Ce^{III}PO_4$ reference [28] has two distinct sets of doublets: $v^0$ at 880.6 eV and $u^0$ at 897.9 eV as a result of Ce $3d_{5/2}$ $4f^2$ ejection, and v' at 884.5 eV and u' at 900.8 eV as a result of the Ce $3d_{5/2}$ $4f^1$ final state, neither of which register a meaningful peak area in this data. Furthermore, from the peak fitting performed (Supporting Information Figure 5), no characteristic band feature of Ce$^{III}$ was found for samples across all crystallite sizes, corroborating the XANES findings that only tetravalent cerium is present.

**Size dependence of the lattice parameter**

The unit cell parameter variation as a function of the NPs' size has been reported for several oxide NPs, including ThO$_2$, CeO$_2$, MgO, Co$_3$O$_4$, Fe$_3$O$_4$, TiO$_2$, *etc.*, but remains a subject of discussion.[2,29–35]

Figure 3 presents the size dependence of the lattice parameters of the as-synthesized CeO$_2$ NPs. When the particle size changes from 2.0 to 91 nm, the unit cell value varies from 5.456 (3) to 5.411 (1) Å. Note that the latter value corresponds to bulk CeO$_2$. In our work and as already reported for CeO$_2$-based NPs,[32,36,2,31] noticeable deviation of the lattice parameter can be observed for particles which sizes are smaller than *ca.* 5 nm. The collected data can be adequately fitted using a power-law relation proposed by Baranchikov *et al.*.[31]

<Figure 3 about here>

According to the paper of Diehm *et al.*,[37] which gathers a large body of experimental data and theoretical calculation, two main models have been proposed to explain this lattice parameter variation. Tsunekawa *et al.* argued that the unit cell is affected by the formation of oxygen vacancies in the smaller NPs, which leads to a change in the oxidation state of the constituting cation. The second model, which is most commonly admitted, attributes the variation to the surface stress resulting from the difference of coordination between atoms on the surface and in the bulk [9,10]. This effect becomes more pronounced as the particle size reduces, *i.e.* as the contribution of the surface atoms to the structural characteristics increases. From our HERFD-XANES and XPS findings, the presence of Ce[III] and hence of the oxygen vacancy, have been discarded, which means that the observed lattice expansion might then be only due to surface stress. To corroborate this assumption, the formation of species present at the NPs surface has been studied by FTIR and TGA measurements. Each FTIR spectrum (

Figure 4) shows an absorption band at *ca.* 450 cm$^{-1}$ characteristic of the $CeO_2$ stretching vibration. Some bands are also visible in the 1300-1700 cm$^{-1}$ region and may be assigned to C-H bending and stretching of C-O bond. The latter may be caused by absorbed $CO_2$ or contamination with ethanol during the synthesis procedure. The signal at 1630 cm$^{-1}$ is associated with the bending frequency of molecular $H_2O$ (H-O-H). The broad adsorption band at *ca.* 3400 cm$^{-1}$ corresponds to hydrated and physically adsorbed water in the sample.[9,38,39] Recently, thermogravimetric analysis coupled with mass spectrometry measurement, performed on $ThO_2$ NPs, have shown that $H_2O$ and $CO_2$ were adsorbed on the NPs surfaces. The corresponding molar fractions of $H_2O$ and $CO_2$ were inversely proportional to the NPs size.[40] Similar observation was obtained from our TGA data presented in the

Figure 4. Ce_2 and Ce_59 have similar thermal decompositions as $Ce(OH)_4$ and $CeO_2$ respectively, while Ce_5.6 and Ce_12.2 exhibit intermediates weight losses. The thermal decompositions of these later is clearly composed of three-well defined steps. The first decomposition step (step I) is observed at 25−180 °C. The second weight loss (step II) occurs at 180−350 °C. The third decomposition step (step III) occurs at 350−700 °C. Our TGA data shows that the weight loss is proportional to the surface to volume ratio, suggesting a higher concentration of adsorbed species on the smaller NPs.

<Figure 4 about here>

The Raman spectrum of bulk $CeO_2$ shows one Raman active fundamental mode at 464 cm$^{-1}$ (Figure 5). This band corresponds to the triply degenerate $T_{2g}$ Raman active mode of the $O_h$ point group and is due to the stretching of the O−O bond. Comparison with the $CeO_2$ NPs reveals a red-shift and an asymmetrical broadening if the $T_{2g}$ mode with decreasing NPs size. Both $T_{2g}$ position and width were fitted and plotted in Figure 5. It can be seen that an increase of the NPs size induces an increase of the $T_{2g}$ position and a decrease of the $T_{2g}$ width. Similar observations has already been reported for $CeO_2$ and other nanocrystals.[41–43] Several factors can contribute to the changes in the Raman peak position and linewidth of the $T_{2g}$ peak with NP size. These include phonon confinement, strain, broadening associated with the size distribution, defects and surface effect.[41–46] In most studies, the red shift is majorly attributed to the lattice expansion and associated strain that occurs when oxygen vacancies are created which leads to the reduction of $Ce^{IV}$ (ionic radius 0.970 Å[47]) in $Ce^{III}$ (ionic radius 1.143 Å[47]). However, the creation of O vacancies and the reduction of $Ce^{IV}$ in $Ce^{III}$ induce a local symmetry distortion and new Raman bands located around 550 and 595 cm$^{-1}$ are observed [48]. The absence of both new Raman bands and detection of $Ce^{III}$ from our HERFD-XANES measurements allows us to discard this hypothesis for the $T_{2g}$ red-shift. One possible explanation could be that the stress generated by the presence of the adsorbed species at the NPs surface that would enhance the downward shift and broadening of the $T_{2g}$.

<Figure 5 about here>

**Size dependence of the electronic structure**

Bulk $CeO_2$ exhibits three characteristic features A, B and C with a doublet structure for B and C leading to a total of 5 bands indicated by A, $B_1$, $B_2$, $C_1$ and $C_2$ in Figure 1. The pre-edge peak (noted A) originates from the dipole-forbidden 2p transition to a mixed 5d-4f valence state while the B and C arises from transitions from $2p_{3/2} \rightarrow 5d_{5/2}$ orbitals. The splitting of $B_1$ and $C_1$ into a doublet structure is due to the crystal field splitting of 5d orbitals.[49] These edge features have been assigned to screened ($B_1$ and $B_2$) and unscreened ($C_1$ and $C_2$) excited states.[22,50]. $B_1$ and $B_2$ features described the $2p \rightarrow 5d$ transition with $4f^1L$ configuration while $C_1$ and $C_2$ are representatives of the $2p \rightarrow 5d$ transition with $4f^0L$ configuration, where L corresponds to the orbital angular momentum.[23,51] In the $CeO_2$ fluorite structure, each Ce atom is surrounded by 8 oxygen atoms located at the corners of a cube creating a cubic crystal field belonging to the $O_h$ point symmetry group. Due to this cubic crystal field, the $Ce^{IV}$ $5d^0$ configuration is split into the $e_g$ and $t_{2g}$ bands corresponding to $B_1$, $C_2$ and $B_2$, $C_1$ respectively.[23,52] The $Ce^{IV}$ valence corresponds to a $5d^0$ configuration implying that in the HERFD-XANES process $t_{2g}$ is firstly filled with electrons while the $e_g$ is empty as the transferred energy is not sufficient. Here the experimental crystal-field energy splitting of Ce 5d in bulk $CeO_2$ is *ca*. 4 eV. This value of energy gap between $e_g$ and $t_{2g}$ is in good agreement with previously published values.[53–55]

Now, looking at the experimental HERFD-XANES spectra of $CeO_2$ NPs, one can observe that all the A, $B_1$, $B_2$, $C_1$ and $C_2$ feature are presented at the same energy position as in bulk $CeO_2$. A value of *ca.* 4 eV is found for the experimental crystal-field energy splitting of Ce 5d in bulk $CeO_2$ NPs, indicating that this energy gap is not affected by the particle size. However, one can note that the $e_g$ feature intensity remains constant for both NPs and bulk $CeO_2$ while the $t_{2g}$ intensity is proportional to the particle size. This experimental observation is corroborated by our theoretical calculations showing that the simulated HERFD-XANES spectra of 2 nm $CeO_2$ exhibit a $t_{2g}$ intensity smaller than that of bulk $CeO_2$. The stability of the $e_g$ feature is also well reproduced. One possible explanation lies in the presence of adsorbed species at the NPs surfaces. In the case of bulk $CeO_2$, the corresponding HERFD-XANES spectra is a direct measurement of the electronic structure of bulk Ce atoms. However, in the case of our NPs, the measured spectra correspond to the average electronic structure of both Ce atoms in the bulk and at the surface. This implies that the observed $t_{2g}$ variation originates from the electronic structure of surface Ce atoms. We showed in the previous section that Ce remains in the IV oxidation state and that the fraction of adsorbed species is inversely proportional to the particle size. In other word, their content is increasing with the surface Ce atoms to bulk Ce atoms ratio. Considering that the $e_g$ level is empty in Ce $5d^0$, the bonding between the surface Ce atoms and the adsorbed species requires the delocalization of the $t_{2g}$ electrons, hence explaining the observed decrease of the $t_{2g}$ intensity on the Ce $L_{III}$ HERFD-XANES spectra. One can also assume that this bonding affects the crystal field by creating new d levels. This larger degeneracy of the $t_{2g}$ level is clearly observable for the sample Ce_5.6 which exhibits a broader $t_{2g}$ feature (Figure 1).

<Figure 6 about here>

**CONCLUSION**

In this work, we have synthesized fluorite $Ce^{IV}O_2$ NPs and studied the effect of the NP size on both local and electronic structure. By analogy with other metal oxides, we have shown that the lattice parameter expands with decreasing particle size. The presence of mainly $Ce^{IV}$, demonstrated by XPS and HERFD-XANES, indicates that the unit cell size-dependence is not linked to the Ce valence but to surface stress. Indeed, our TGA and FTIR data confirms the presence of surface hydroxyl and carbonate groups that have a tensile effect on the crystalline lattice. Additionally, the size-dependence of the electronic structure, and especially of the $t_{2g}$ feature in the Ce $L_{III}$ XANES spectrum, have been experimentally evidenced and confirmed with theoretical calculations.

FIGURES

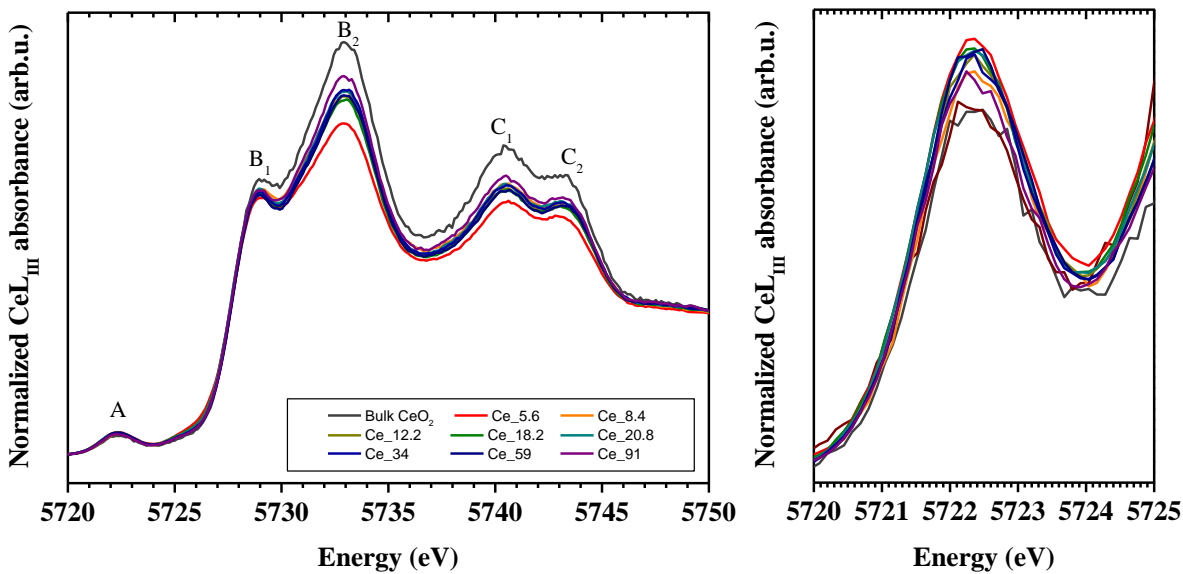

Figure 1: (Left) Ce L$_3$ HERFD-XANES spectra of CeO$_2$ NPs with different sizes compared to bulk CeO$_2$ (> 100 nm). (Right) Pre-edge region of the Ce L$_3$ HERFD-XANES spectra.

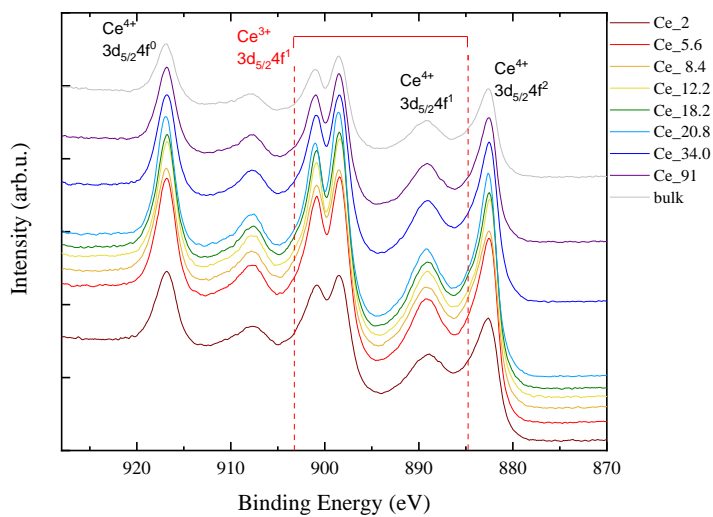

Figure 2: High-energy resolution photoemission spectra of the Ce 3d spectral region. The spectra confirm the presence of the dominating Ce$^{IV}$.

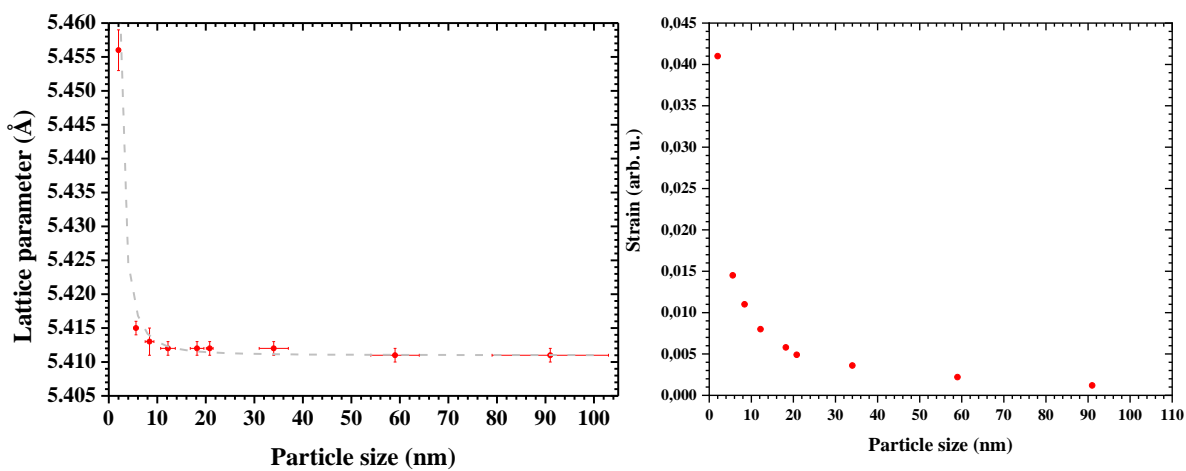

Figure 3: Decrease of both lattice parameter (left) and strain (right) with the particle size. The lattice parameter, the strain and the particle size have been calculated from the Rietveld refinement of the corresponding XRD patterns. All the values are provided in the Table 1 of the SI. The gray dotted line corresponds to the power-law relation proposed by Baranchikov et al.[31].

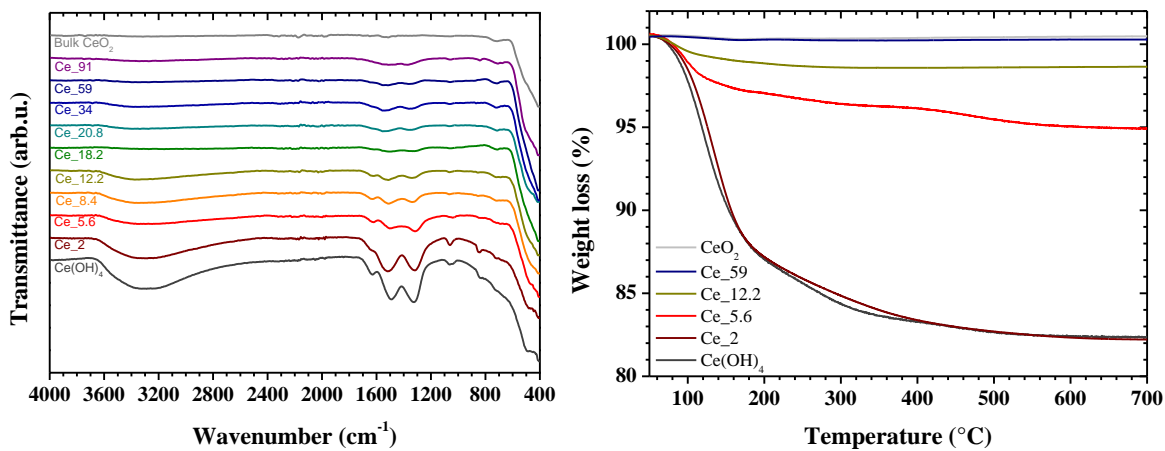

Figure 4: Determination of the adsorbed surface species. (Left) FTIR spectra of $CeO_2$ NPs compared with $Ce(OH)_4$ and bulk-$CeO_2$ references; (Right) Thermogravimetric analysis data of $CeO_2$ NPs of different sizes compared with $Ce(OH)_4$ and bulk-$CeO_2$.

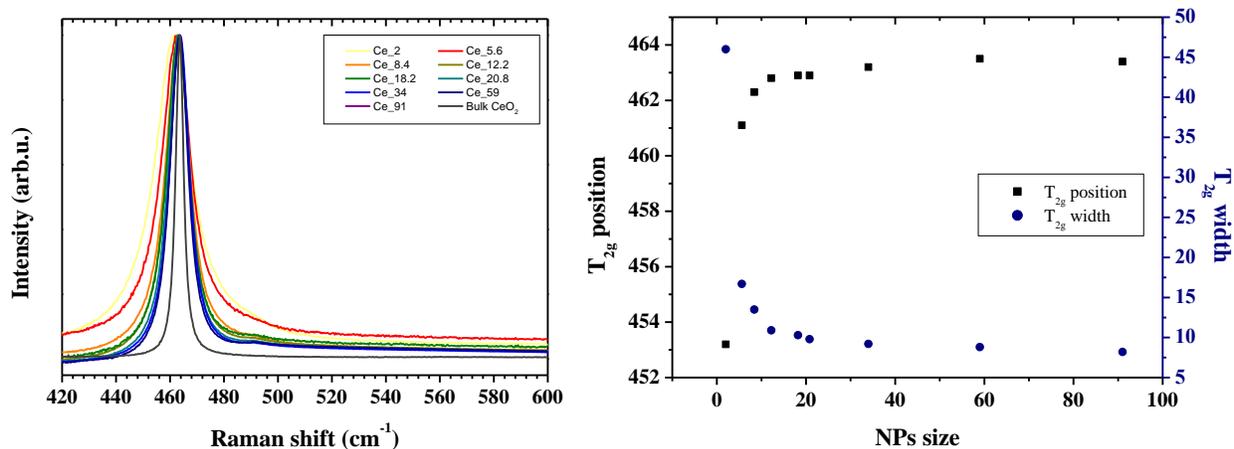

Figure 5: (Left) Raman spectra (Right) T$_{2g}$ position and width as a function of the particle size

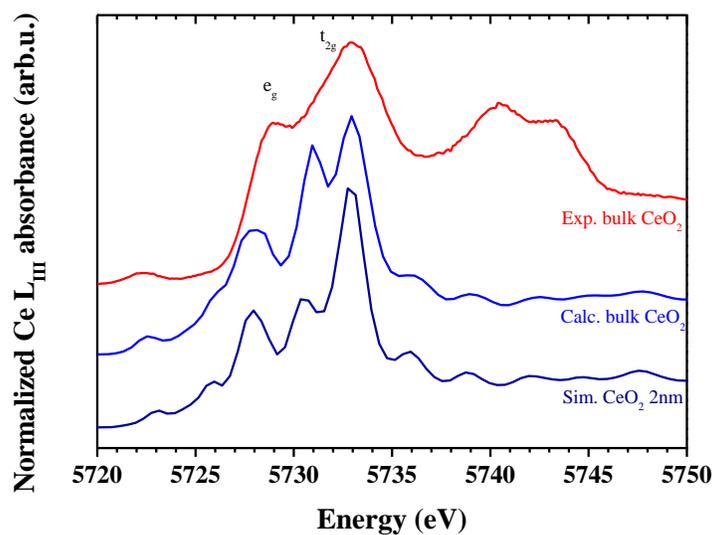

Figure 6: Comparison of experimental and calculated Ce L$_{III}$ HERFD-XANES spectra of both 2 nm CeO$_2$ and bulk CeO$_2$. Calculated data shown here are reproduced from T.Plakhova *et al.*[21]. Theoretical calculations predicts a lower t$_{2g}$ intensity for the 2 nm CeO$_2$.

TABLES

Table 1: Synthesis route, particle size and lattice parameter. Note that Ce_2 was measured by HERFD-XANES but the results are not presented here due to the beam damage.

| Sample Name | Synthesis route | Crystallite size (nm)* | Lattice parameter (Å)* | Strain (%) * |
|---|---|---|---|---|
| Ce_2 | Hydrothermal treatment of Ce(IV) hydroxide (3h at 150°C under autogenic pressure) | 2.0 ± 0.1 | 5.456(3) | 0.041 |
| Ce_5.6 | Hydrothermal treatment of Ce(IV) hydroxide (3h at 200°C under autogenic pressure) | 5.6 ± 0.4 | 5.415(1) | 0.015 |
| Ce_8.4 | Ce_2 heated 1h at 350°C in air | 8.4 ± 0.9 | 5.413(2) | 0.011 |
| Ce_12.2 | Ce_4.2** heated 1h at 500°C in air | 12.2 ± 1.5 | 5.412(1) | 0.008 |
| Ce_18.2 | Ce_4.2** heated 1h at 700°C in air | 18.2 ± 1.4 | 5.412(1) | 0.006 |
| Ce_20.8 | Ce_4.2** heated 1h at 800°C in air | 20.8 ± 0.7 | 5.412(1) | 0.005 |
| Ce_34 | Ce_4.2** heated 1h at 900°C in air | 34 ± 3 | 5.412(1) | 0.004 |
| Ce_59 | Ce_4.2** heated 1h at 925°C in air | 59 ± 5 | 5.411(1) | 0.002 |
| Ce_91 | Ce_4.2** heated 1h at 950°C in air | 91 ± 12 | 5.411(1) | 0.001 |

\* derived from the refinement of the XRD patterns.
\*\* Ce_4.2 nanopowder was synthesized by hydrothermal treatment of Ce(IV) hydroxide (3h at 180°C under autogenic pressure). The particle size of Ce_4.2 was estimated to be 4.2 ± 0.3 nm from the XRD pattern.

ASSOCIATED CONTENT

(Word Style "TE_Supporting_Information"). **Supporting Information**. A listing of the contents of each file supplied as Supporting Information should be included. For instructions on what should be included in the Supporting Information as well as how to prepare this material for publications, refer to the journal's Instructions for Authors.

The following files are available free of charge.

brief description (file type, i.e., PDF)

brief description (file type, i.e., PDF)

AUTHOR INFORMATION

**Corresponding Authors**

*E-mail: d.prieur@hzdr.de; E-mail : walter.bonani@ec.europa.eu

**Author Contributions**

The manuscript was written through contributions of all authors. All authors have given approval to the final version of the manuscript. ‡These authors contributed equally. (match statement to author names with a symbol)

**ACKNOWLEDGMENT**

D. P., K.P. and W.B. acknowledge the KIT light source for provision of instruments at their beamlines and the Institute for Beam Physics and Technology (IBPT) for the operation of the storage ring, the Karlsruhe Research Accelerator (KARA). D.P., K.P. and W.B. are thankful to Dr. Kathy Dardenne and Dr. Joerg Rothe for their help during the beamtime. D.P. is also indebted to the ID26 team to lend us the crystals for XAS measurements. In addition, K.P., W.B. and O.W. acknowledge Sarah Stohr and Herwin Hein for the technical support and thermogravimetric analysis and Dr. Heike Störmer from the Laboratory for Electron Microscopy of the Karlsruhe Institute of Technology (KIT) for TEM images.

X.G. acknowledges the institutional funds from the Department of Chemistry at Washington State University. K.W.K. acknowledges the support through the WSU Radiochemistry Traineeship. A portion of this research was performed using EMSL (grid.436923.9), a DOE Office of Science User Facility sponsored by the Office of Biological and Environmental Research.

K.O.K acknowledges support from European Research Council (grant N 75969).

BRIEFS (Word Style "BH_Briefs"). If you are submitting your paper to a journal that requires a brief, provide a one-sentence synopsis for inclusion in the Table of Contents.

SYNOPSIS (Word Style "SN_Synopsis_TOC"). If you are submitting your paper to a journal that requires a synopsis, see the journal's Instructions for Authors for details.